\documentstyle[prb,aps,epsf]{revtex}
\input epsf
\begin{document}
% \draft command makes pacs numbers print
\draft
%\twocolumn[
%\hsize\textwidth\columnwidth\hsize\csname @twocolumnfalse\endcsname
% \twocolumn

\title{ Superconductivity in mesoscopic high-$T_{c}$ superconducting particles }
\author{V.~A.~Ivanov\cite{A1}, V.~R.~Misko\cite{A2}, V.~M.~Fomin\cite{A3},
J.~T.~Devreese\cite{A4}}
\address{Theoretische Fysica van de Vaste Stoffen, 
Universiteit Antwerpen (U.I.A.), \\
Universiteitsplein 1, B-2610 Antwerpen, Belgi\"e}
\date{\today}
\maketitle

\begin{abstract}
Based on the Hubbard model in the framework of non-phonon kinematical mechanism 
and taking into account the discreetness of an electronic energy spectrum, 
the superconducting critical temperature of a mesoscopic high-$T_{c}$ sphere
is analyzed as a function of doping and as a function of a particle's radius. 
The critical temperature $T_{c}$ is found to be an oscillating function of the 
radius of a particle. 
The size-dependent doping regime is revealed in high-$T_{c}$ nanoparticles. 
Our analysis shows that each oscillation in $T_{c}$ corresponds to the increase 
of a number of the energy levels in the sphere by one.
The amplitude of oscillations of $T_{c}$ increases with decreasing $R$ and 
can reach a value of 6~K for nanopartilces with sizes about 25~nm,
in a good agreement with experimental studies of YBa$_{2}$Cu$_{3}$O$_{7-\delta}$
nanoparticles. 

\end{abstract}

\section{Introduction}

After experiments by Ralph {\it et al.} \cite{Ralph95,Ralph97} on the
electronic states in nanosize {\it Al} particles there has been increasing
experimental and theoretical interest in properties of ultrasmall
superconducting samples with discrete fermionic energy spectrum (see \cite
{Delft1,Delft2} and Refs therein). Incorporation of the nanosize particles
as a part of single-electron tunneling transistor schemes allowed for
measurements of excitation spectra as a function of the number of electrons
in a superconducting particle. The large spectroscopic gap between the
lowest energy level and all the others which can be driven to zero by
application of magnetic fields has been explained by superconducting pairing.

Till now experiments have been completed for the low {\it T}$_{c}$ elemental
superconductors (Al, Sn, ...), showing traces of superconducting
pairing for metallic particles even with the electron energy spacing being
larger than the superconducting energy gap: $d > \Delta$. 
For elemental superconductors in the order of magnitude 
$\Delta \sim T_{c}$, the energy spacing is $d \sim \hbar^{2}/(m_{e}R^{2})$ 
and for these nanoparticles their size 
$R \lesssim \hbar / \sqrt{m_{e}T_{c}}$. 
Thus, to satisfy the regime $d \gtrsim \Delta$ 
for characteristic critical temperature values $T_{c}\sim $1K the 
nanoparticle size should not exceed $\sim $ 30~nm \cite{murayama}. 
As shown by Tinkham team, the problems of the sample preparation with these
parameters have been solved by modern technology. 
In such nanoparticles from
elemental superconductors their average size is smaller or comparable with
the coherence length and fluctuations of superconducting order parameter
do not allow the mean-field treatment. On this way the recent theoretical work
includes studies of mesoscopic systems with equally spaced levels 
\cite{vonDelft}, Wigner-Dyson level spacing \cite{Smith}, parity effect \cite{Matveev}, 
fluctuation effects \cite{Smith1}, fixed-N canonical treatment \cite{Broun}
and the suppressing role of the non-pair portion of the interaction \cite{Crespi}. 
The differences in energy level spectra for even and odd numbers
of electrons are being interpreted as a result of superconducting paring
interactions.

But today the nanotechnology is on the threshold of production of nanosize
samples from high-T$_{c}$ cuprates \cite{geohegan,mozhaev,zhao}.
More recently, a several degrees enhancement of the onset transition
temperature $T_{c}$ has been observed in superconducting dots formed by the
high-$T_{c}$ superconductor YBa$_{2}$Cu$_{3}$O$_{7-\delta }$ (YBCO) \cite{menon} 
and in YBa$_{2}$Cu$_{3}$O$_{7-\delta }$\ films with detected
spherical particles on their surface \cite{mozhaev}. The dots were formed
when powder YBCO particles with a diameter $\leq 600$~nm were exposed to a RF
plasma to produce a Coulomb crystal. The plasma damaged the particles and
caused 20-25~nm sized isolated islands of the correct stoichiometry to
segregate within each particle. Possible technological applications of
superconducting nanoparticles make it desirable, to engineer their energy
spectra and the pairing characteristics. The crucial question that has
arisen in this connection is: how can the superconducting critical
temperature for a high-$T_{c}$ nanoparticle be controlled by changing its
size and the doping degree?

In nanoparticles made from high-$T_{c}$ materials it's questionable to reach
the regime $d > \Delta$ corresponding to size $R \lesssim \hbar/\sqrt{m \Delta}=
\hbar\sqrt{2}/\sqrt{mT_{c}(2\Delta /T_{c})}$. 
Really for high-$T_{c}$ superconductors the BCS ratios 2$\Delta /T_{c}$ are quite 
large (9 for La$_{2-x}$Sr$_{x}$CuO$_{4}$, 6-8 for YBa$_{2}$Cu$_{3}$O$_{7-\delta }$,
and even 12 for Bi$_{2}$Sr$_{2-x}$La$_{x}$CuO$_{6}$ - see \cite{maksimov}
and Refs. therein) and the magnitudes of an effective electron mass $m$
are enhanced due to narrow energy bands of high-$T_{c}$ materials. 
As a result, the high-$T_{c}$ nanoparticle size $R$ is becoming comparable
with a unit cell size (it's lattice constant parameter $c \simeq 1.2$~nm in
YBCO) or even less than the latter with a significant deviations of a
chemical composition and the intra-size vibrations from those in a bulk
mother substance. 
Physically, it means that for the nanosize particles
formed from stochiometric high-T$_{c}$ material we can not reach regime 
$\Delta < d$.
Thus, the field of the nano-size high-$T_{c}$ superconductors is
renewing the earlier Anderson idea \cite{anderson} about the disappearance of
superconductivity in particles, which are so small that the ``granularity''
of an electron energy $d$ is larger than the superconducting energy gap $\Delta $. 
So, for the nanosized high-T$_{c}$ particles we are returning back to the limit 
$d < \Delta $.

Long time ago the studies of the superconducting-transion temperature $T_{c}$ 
of metallic films of Al, Sn, In, Zn prepared by vacuum
deposition onto substrate held at cryogenic temperatures revealed the quantum
size effect of the $T_{c}$-oscillations with a film thickness
\cite{zavaritzky,khukhareva,strongin,abeles,komnik,strongin1,orr}
(see also review \cite{tavger}). 
In the simplest approaches \cite{thomson,paskin,kirzhnitz} 
the quantum size effects follow from oscillations in the
electronic density of states with thickness which in turn lead to
oscillations with thickness of $T_{c}$ and other film quantities.

Here in the Anderson regime $d < \Delta $ we consider a similar mechanism of finite-size effects aiming the goal to apply it to realistic high-$T_{c}$ nano-particles. 
Worthy to note, that an evaluated size of nanoparticles under consideration (see next
section) exceeds the small coherence length scale with neglect of
fluctuations and the Aslamazov-Larkin effect \cite{aslamazov}. 
The method is
based on the study of the non-phonon kinematical superconductivity (see \cite{ivan_el} 
and references therein) in systems with strongly interacting
electrons and an intrinsic ``granularity'' of energy, induced by the
discreteness of electron energy levels in a nanoparticle. 
Formally the non-phonon mechanism allows to avoid difficulties caused by phonon cut-off
energies in small particles. 
In the model we assume the boundary condition on the electronic wave-functions is that they vanish on the surface of spherical nanoparticle and we neglect the surface effects.

\section{The model}

In the present communication we tackle the problem starting from the
strongly interacting limit of the Hubbard model ($U \gg t$), 
$H=-\sum t_{ij}X_{i}^{\sigma 0}X_{j}^{0\sigma }-\mu \sum n_{i}$, for
correlated electrons ($\mu $: their chemical potential, $n_{i}$: electron
density operator on site, $t_{ij}$: intersite electron hopping) in the YBCO
nanoparticles of a spherical shape. 
As usual, $X^{0\sigma }$ denote a
projection Hubbard operator with an index ``$0\sigma $'' labeling an
intrasite transition from an energy level occupied by an electron with spin
projection $\sigma $ to an empty one: $\sigma \rightarrow 0$. 
In high-$T_{c}$ materials the characteristic strong electron interactions remove
the orbital degeneracy of energy levels. 
The electron energies have been
extracted \cite{ivan_el,ivan_jp} from zeros of the inverse Green's function,
obeying the Dyson equation 
\begin{equation}
{\mathcal{D}}^{-1}(i\omega )=[{\mathcal{D}}^{0}(i\omega )]^{-1}+t_{ij},  \label{dyson}
\end{equation}
where the first order self-energy is itself an electron hopping $t_{ij}$
inside a spherical nanoparticle. The zeroth order Green's function is
defined as ${\mathcal{D}}^{0}(\tau )=-\langle {\hat{T}}X^{0\sigma }(\tau
)X^{\sigma 0}(0)\rangle $ or in Matsubara $\omega =(2n+1)\pi T$
-representation it is such as 
\begin{equation}
{\mathcal{D}}^{0}(i\omega )=\frac{f}{-i\omega _{n}-\mu }.  \label{gfzow}
\end{equation}
The correlation factor, 
\begin{eqnarray}
f &=&\langle X^{00}+X^{\sigma \sigma }\rangle = 
1-\langle 
X^{ {\overline{\sigma }}{\overline{\sigma }} }
\rangle  
\label{corfac}
\\
&\equiv &1-n_{{\overline{\sigma }}}=1-\frac{n}{2}=\frac{1+x}{2}, 
\nonumber
\end{eqnarray}
is governed by an average electron density per site $n$ or/and$\ $by a
doping parameter $x=1-n$ ($0\leq x\leq 1$).

In the momentum representation of bulk material the correlated electron
energy is $\xi =ft_{p}-\mu $ with an energy dispersion values $t_{p}$
located inside the band: $-w\leq t_{p}\leq w$. The superconducting
instability is driven by the homogeneous Bethe-Salpeter equation for a
vertex $\Gamma $ of two scattering fermions with opposite spin orientations
and opposite momenta in their reference system: 
\begin{equation}
\Gamma =2T\sum\limits_{p,n}t_{p}G_{-\omega }(-p)G_{\omega }(p)\Gamma (p),
\label{bs}
\end{equation}
where $G_{\omega }(p)=\left( -i\omega +\xi _{p}\right) ^{-1}$ is the normal
Green's function. After summation over Matsubara frequencies this equation
can be converted to equation

\begin{equation}
1=\sum\limits_{p}\frac{t_{p}}{ft_{p}-\mu }\mbox{th}\frac{ft_{p}-\mu }{2T_{c}}%
,  \label{bsp}
\end{equation}
represented in momentum representation.

In the experiment \cite{menon} both the initial YBCO material and nanosized 
(20-25 nm) islands produced by plasma damage are polycrystalline particles 
with chaotic distribution of crystallographic directions. 
Therefore, we can model each nanosized island by isotropic sphere of 
radius $R$. 
As in the case of dirty superconductors \cite{anderson}, in a spherical
particle of radius $R$ with the quantized electronic energy levels, 
$E_{\nu}=\hbar ^{2}\nu ^{2}/\left( 2mR^{2}\right)$, the momenta are not good
quantum numbers. In the absence of non-elastic processes the suitable
quantum numbers are the eigen-energies of the problem, shifted by a
half-bandwidth $w$: $\epsilon _{\nu }=E_{\nu }-w$ ($-w\leq \epsilon _{\nu
}\leq w$). For HTSC's of interest such as YBa$_{2}$Cu$_{3}$O$_{7-\delta }$ a
half-bandwidth parameter has characteristic magnitude $w \sim 0.5$~eV 
\cite{radtke}. The lowest energy level is at the bottom of the starting bulk
energy band, $\epsilon _{0}=-w$, whereas the highest one, $\epsilon _{\nu
_{0}}$, coincides with the top of the band: $\nu _{0}=\left[R\sqrt{2mw}/\hbar\right] 
$ ($\left[ ...\right] $\ stands for an integer number). So, in an
energy representation the superconducting critical temperature $T_{c}$ is
determined by the following equation: 
\begin{equation}
1=\frac{1}{\nu _{0}}\sum_{\nu =1}^{\nu _{0}}\frac{\frac{E_{\nu }}{w}-1}
{\left( \frac{E_{\nu }}{w}-1\right)f -\frac{\mu }{w}}\tanh \left( 
\frac{\left( \frac{E_{\nu }}{w}-1\right)f -\frac{\mu }{w}}{\frac{2T_{c}}{w}}
\right),  
\label{ct}
\end{equation}
where the summation is running over quantized energy levels till the upper
level $\nu _{0}$; the
chemical potential $\mu $\ has a meaning of the highest occupied energy
level. The doping parameter $x$ is defined by 
\begin{equation}
1-x=2T\sum\limits_{n,\nu = 1}^{\nu _{0}}e^{i\omega_{n} \delta }{\mathcal{D}}(\nu
,\omega_{n} ).  \label{xmud}
\end{equation}
From here one can find the equation for the chemical potential $\mu $: 
\begin{equation}
\frac{1-x}{1+x}=\frac{R}{%
%TCIMACRO{\UNICODE[m]{0x127}}%
%BeginExpansion
h\hskip-.2em\llap{\protect\rule[1.1ex]{.325em}{.1ex}}\hskip.2em%
%EndExpansion
\nu _{0}}\sqrt{\frac{m}{2}\left( \frac{2\mu }{1+x}+w\right) }.  \label{xmu}
\end{equation}
The system of equations, Eqs.~(\ref{ct}), (\ref{xmu}) should be solved
self-consistently to obtain $T_{c}$.

The applied arithmetics is valid until the superconducting pairing is
allowed for electrons belonging to different energy levels in a
nanoparticle, {\it i.e.} the thermal energy should exceed the distance
between the discrete energy levels, namely $T_{c}>$%
%TCIMACRO{\UNICODE{0x127}}%
%BeginExpansion
h\hskip-.2em\llap{\protect\rule[1.1ex]{.325em}{.1ex}}\hskip.2em%
%EndExpansion
$^{2}\left[ \nu _{0}^{2}-\left( \nu _{0}-1\right) ^{2}\right] /(2mR^{2})$,
wherefrom it follows the restriction for the high-$T_{c}$\ nanoparticle
radii in our approach: 
\[
R>R_{0}=\frac{2\hbar}{T_{c}}\sqrt{\frac{w}{m}}. 
\]
For estimate we assume the effective mass enhancement $m/m_{e}\approx 10$ 
from the earlier optical reflectivity measurements \cite{thomas} 
for the YBCO bulk crystals. Noteworthy, the more recent data for
YBCO \cite{basov} give closed estimate $m\approx 8-9m_{e}$ under the
assumption of a twice band mass enhancement due to electron correlations: 
$m_{b}=2m_{e}$. 
Such estimate $m \approx 10m_{e}$ is also in an agreement with
theoretical evaluations \cite{renato,alexandrov} for an effective
electron mass in YBCO. Finally, for the YBCO nanoparticle of interest \cite{menon} 
with $T_{c}=92$~K and a half-bandwidth $w=0.725$~eV 
(this value of $w$ corresponds in our calculations to the ``bulk'' critical temperature 
$T_{c}=92$~K)
one can
conclude {\it the lower critical radius of a spherical YBCO nanoparticle
such as} $R_{0}=15$~nm to apply the theory. 
The nanoparticle with a critical
radius still contains 75$\pi \cdot 10^{3}$ unit cells of YBCO (the YBCO 
unit cell volume is $\sim 0.18$~nm$^{3}$), which is in a reasonable
agreement with a neglect of surface and fluctuation effects. 
As it will
be seen from numerical results in next sections, although we are working in
the bulk superconducting regime, the quantum size effects due to a
''granularity'' of an energy in high-T$_{c}$ nanosize particles are still
traced.

\section{Results and discussion}

\subsection{The critical temperature $T_{c}$ of a mesoscopic high-$T_{c}$
superconducting sphere as a function of doping $x$}

Based on Eqs.~(\ref{ct}) and (\ref{xmu}), we analyze the critical
temperature of a mesoscopic high-$T_{c}$ superconducting sphere as a
function of doping. 
We have performed calculations for various radii $R$ of a superconducting sphere
varying from $R=15$~nm to 1000~nm. 
In Fig.~1, the critical temperature $T_{c}$ of a mesoscopic high-$T_{c}$ 
superconducting sphere is plotted as a function of doping $x$ for 
particles with radii $R=15$~nm, 16~nm, 20~nm, 30~nm and 1000~nm. 
For any radius within the above region, $T_{c}$ as a function of $x$ 
is characterized by the following behavior: 
$T_{c}$ is zero until some minimal doping $x_{min}$, which provides a particle 
to be superconducting; 
then $T_{c}$ increases (underdoped regime) and reaches a maximum at some 
value of doping (optimal doping); 
further increase of $x$ leads to decrease of $T_{c}$ (overdoped regime) (see Fig.~1). 

\begin{figure}
\protect\centerline{\epsfbox{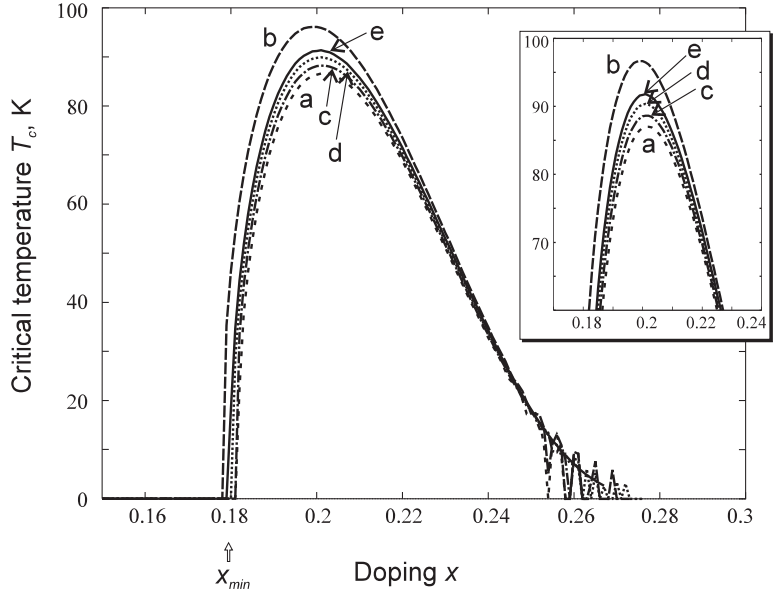}}
\smallskip 
\caption{
The critical temperature $T_{c}$ as a function of doping $x$, for a mesoscopic 
high-$T_{c}$ superconducting spherical particle. 
The radii of the particles are 
$R=15$~nm (a), 16~nm (b), 20~nm (c), 30~nm (d) and 1000~nm (e). 
The value of the bandwidth parameter for YBa$_{2}$Cu$_{3}$O$_{7-\delta}$ is 
$w=0.725$~eV.
The inset shows upper parts of the curves $T_{c}=T_{c}(x)$ for temperatures ranging in 
the interval from 60 to 100~K.
}
\label{fig1}
\end{figure}

The maximal value of $T_{c}=T_{c}(x)$ is a non-monotonous function of $R$. 
For example, at $R=15$~nm the maximum of $T_{c}$ is 86.6~K, 
whereas at $R=16$~nm the maximumal value of $T_{c}$ is as high as 96.1~K.
With increasing radius, a deviation in the maximal value of $T_{c}$ reduces, and 
for large radii (more than 100~nm) the maximal values of $T_{c}$ are close to each other 
and lie in the vicinity to $T_{c}\approx 92$~K. 
A further increase of the radius of the sphere does not
lead to any appreciable changes of $T_{c}$.

A position of the maximum of $T_{c}=T_{c}(x)$ itself also varies as a function 
of radius $R$. 
For $R=15$~nm, the maximum of $T_{c}=T_{c}(x)$ is reached at a higher $x$ 
as compared to ``bulk'' value ($R=1000$~nm). 
On the contrary, for $R=16$~nm, the maximum of $T_{c}=T_{c}(x)$ is 
at a lower $x$ as compared to the ``bulk'' value. 

The critical temperature of a mesoscopic high-$T_{c}$ superconducting sphere
as a function of doping is characterized by some minimal value of doping 
and by oscillating behaviour in strongly overdoped regime (see Fig.~1). 
Although the critical temperature of a mesoscopic high-$T_{c}$ superconducting 
sphere is itself a smooth function of doping, the minimal doping $x_{min}$ occurs 
to be an oscillating function of the radius of the particle. 
The minimal doping, which provides the particle to be superconducting, is
very sensitive to the bandwidth parameter $w$. 
Varying the parameter $w$, we can conclude that 
reduction of the bandwidth leads to increase of the average value of 
$x_{min}$ and to decrease of the period of oscillations.

\subsection{The critical temperature of a mesoscopic high-$T_{c}$
superconducting sphere versus its radius}

In this section, we analyze in detail the oscillations in $T_{c}$ that were
noted in the previous subsection. 
The critical temperature $T_{c}$ is calculated as a function of the radius $R$ 
of the sphere for various values of doping $x$. 
In Fig.~2, $T_{c}=T_{c}(R)$ is shown for $x=0.2$ (that is close to the optimal doping). 
The critical temperature $T_{c}$ as a function of radius $R$ is characterized by 
an oscillating behavior. 
The amplitude of these oscillations decreases with increasing $R$. 
It is maximal at small $R$: $T_{c}=T_{c}(R)$ varies from about 86.4~K up to 97.8~K 
in the interval from $R=15$~nm to 18~nm (see the upper inset to Fig.~2). 
Then, as shown in the bottom inset to Fig.~2, the amplitude of the oscillations 
decreases very fast up to $R \approx 100$~nm. 
For larger $R$, $T_{c}=T_{c}(R)$ is characterized by a slow decrease versus $R$. 
The function $T_{c}=T_{c}(R)$ ``saturates'' with increasing $R$ and reaches the 
``bulk'' value 92~K. 

\begin{figure}
\protect\centerline{\epsfbox{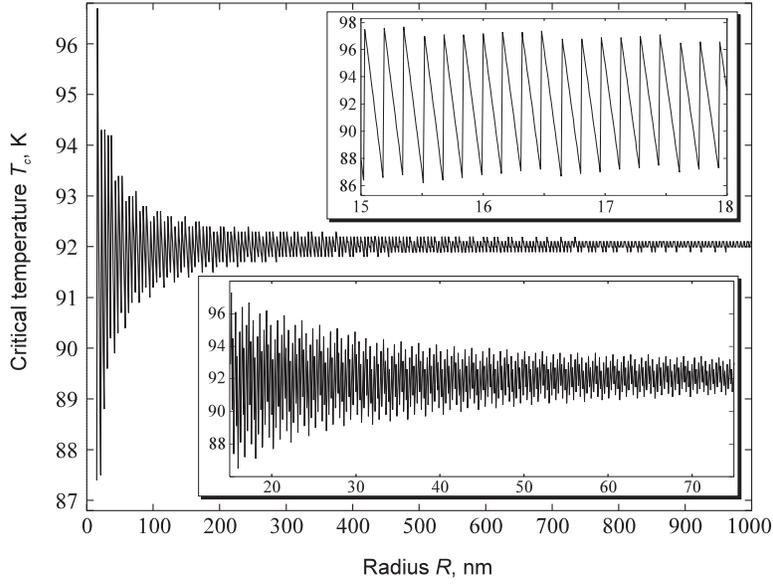}}
\smallskip 
\caption{
The critical temperature $T_{c}$ as a function of the radius $R$ of a 
mesoscopic high-$T_{c}$ superconducting particle, at the optimal doping $x=0.2$, 
for radii from $R=15$~nm to 1000~nm. 
In the bottom inset: $T_{c}$ as a function $R$ for radii from $R=15$~nm to 75~nm. 
The top inset: $T_{c}=T_{c}(R)$ for $R=15$~nm to 18~nm. 
Each oscillation in $T_{c}=T_{c}(R)$ corresponds to the change of the 
number of levels in the system by one.}
\label{fig2}
\end{figure}

The oscillations in the function $T_{c}=T_{c}(R)$ are due to the discrete
density of states in the superconducting sphere. 
Recall that we analyze high-$T_{c}$ spheres with a confinement of carriers 
in all three dimensions.
Our analysis shows that each oscillation in $T_{c}$ with increasing $R$ corresponds
to the increase of a number of the energy levels in the sphere by one.

In the overdoped regime, the function $T_{c}=T_{c}(R)$ is analyzed for $x=0.225$. 
The results of the calculations are shown in Fig.~3 for $R=15$~nm to 1000~nm
The ``bulk'' value of $T_{c}$ for $x=0.225$ ($T_{c,bulk}=62.4$~K) is much 
lower than that for the optimal doping. 
Therefore, slowly varying the doping parameter $x$, we can model various real samples 
used in experiments.

\begin{figure}
\protect\centerline{\epsfbox{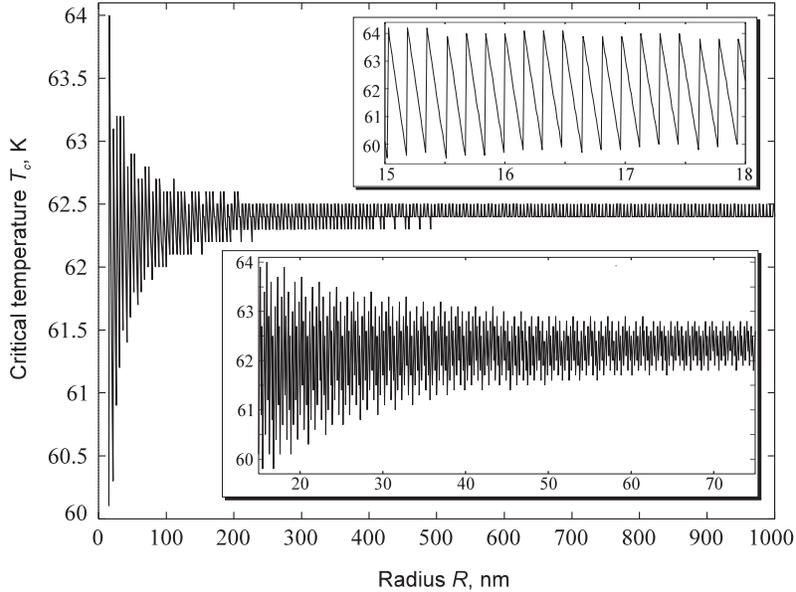}}
\smallskip 
\caption{
The critical temperature $T_{c}$ as a function of $R$ in the overdoped regime, 
$x=0.225$. 
The range of particles radii and the bandwidth parameter are the same as in Fig.~2.}
\label{fig3}
\end{figure}

In the paper by Menon {\it et al.} \cite{menon}, a slight enhancement in the
critical temperature $T_{c}$ was detected for YBa$_{2}$Cu$_{3}$O$_{7-\delta}$ 
plasma-damaged nanoparticles with sizes about 25~nm. 
To find the transition temperature in the particles, their dc susceptibility was measured 
as a function of temperature in a SQUID magnetometer.
The transition temperature  for the onset of superconductivity is defined as the 
temperature at which the susceptibility begins to drop \cite{menon}. 
In some cases, the increase in critical temperature was only 1~K ($T_{c,bulk}=92$~K), 
whereas the increase reached 6~K for some other samples \cite{menon}. 
From our calculations it follows that, for the optimal doping
(characterized by the ``bulk'' value $T_{c,bulk}=92$~K) the enhancement of the critical
temperature for nanoparticles with the radii $R=15$~nm to $R=18$~nm can be up to 6~K 
(see upper inset to Fig.~2).
This is in a rather good agreement with the experimental results \cite{menon}. 
Moreover, the revealed oscillations of $T_{c}$ as a function of $R$ explain the 
non-regular character of the enhancement of $T_{c}$ in different samples \cite{menon}.

\section{Conclusions}

The superconducting critical temperature of a mesoscopic high-$T_{c}$ sphere
has been analyzed as a function of doping and as a function of a
particle's radius. The approach is based on the Hubbard model as the most 
fundamental model of the electronic interactions in high-$T_{c}$ materials. 
Quantum Monte Carlo studies of this model (see \cite{dagotto} and references 
therein) concluded that the enhanced superconducting correlations are not 
predominant at low temperatures or in the ground state. 
Although some of numerical works supported the existence of superconducting 
pairing in the Hubbard model \cite{hirsch,husslein}, these quantum Monte Carlo 
computations have restricted parameter space such as the system size, the doping, 
the low values of the Hubbard energy etc. 
The modern Monte Carlo \cite{giamarchi,yanagisawa,yamaji} 
and other \cite{maier} metods allow one to treat a
wider parameter space with large magnitude of the Hubbard energy and reveal
the superconducting correlations in the ground state of the two-dimensional
Hubbard model. So, the numerical simulation of superconductivity in the
ground state of the Hubbard model still remains the issue for the serious
problem of the two-dimensional nature of cuprate high-$T_{c}$ superconductors. 
As distinct from that, the kinematical mechanism (see Ref.~\cite{ivan_el} 
and references therein) considers the superconducting instability of 
strongly correlated electrons in normal state beyond the knowledge of their 
ground state properties. 
Therefore the present study of superconductivity is based on the
non-phonon kinematical mechanism in the framework of the Hubbard model
taking into account the ``granularity'' of electron energies (the discreteness
of an electronic energy spectrum) in nanoparticles. The chaotic crystallite
structure allowed us to model nanoparticles by the isotropic spherical
particles. This point made suitable the applied three-dimensional scenario
of the kinematical attraction for YBCO nanoparticles under study. Worthy to
note that in one-band two-dimensional Hubbard model the kinematical
superconducting instability occurs at bulk doping $x<1/3$ as in variational
Monte Carlo method (see, e.g., Ref.~\cite{giamarchi}).
Stimulated by the recent successful results on technology 
of a nano-sample preparation from YBCO \cite{geohegan,mozhaev,zhao,menon},
the typical superconducting compound YBa$_{2}$Cu$_{3}$O$_{7-\delta }$ has
been taken as a mother substance forming the nanoparticle. We have shown
that the critical temperature of a mesoscopic high-$T_{c}$ superconducting
sphere is characterized by a non-monotonic dependence on doping as in bulk material. 
However, contrary to the bulk high-$T_{c}$ materials the $T_{c}$-maximum
(corresponding to the optimal doping) is shifted to the lower doping regime.
A bulk material with a decreased $T_{c}$ value in
overdoped regime can form a nanoparticle with a maximal $T_{c}$ value which
means that it is in an effective optimally doped regime. 
Also an optimally
doped bulk high-$T_{c}$\ material behaves as an underdoped substance in
nanoparticles.
The revealed size-dependent doping regime in high-$T_{c}$ nanoparticles 
is new effect in comparison with quantum size effects in elemental 
low-$T_{c}$ nanoparticles.  

Although we worked in the Anderson regime $d < \Delta $, {\it i.e.} neglecting 
the parity effects, and due to the
small coherence length of high-$T_{c}^{\prime}s$, $\xi < R_{0}=15$~nm,
neglecting the fluctuation effects, it have been established that the
critical temperature of a mesoscopic high-$T_{c}$ superconducting sphere is
an oscillating function of a large sphere size $R > R_{0}$. 
For sphere radii
with value $R \gtrsim 1000$~nm and doping $x=0.2-0.25$ the $T_{c}$-oscillations
actually disappear. 
The oscillating quantum size effects appear whenever a
size-dependent energy level passes through the chemical potential as the
size of a nanosphere is varied. 
At each resonance a new quantized energy
level starts to contribute to an electron density of states and
superconducting quantities. 
For chemical potential nearly zero ($x \sim 1/3$) this oscillating behaviour 
is more pronounced and results in new $T_{c}$ oscillations with doping 
in strongly overdoped regime (cf. Fig.~1). 
The amplitude of $T_{c}$-oscillations is enhanced 
for lower radii of mesoscopic spheres. 
For an electron confined
to a sphere of radius $R$, the radial part of the wave function
is proportional to $j_{1}(kR)$, {\it i.e.} to $\sin(kR)/kR$ for electrons without angular momenta. 
Therefore the
boundary condition yields the relationship $kR=n\pi$ with a period of
oscillations such as $\Delta R=\pi /k=\lambda _{F}/2$, a half of de Broglie
wave-length. 
The calculations are performed for spherical YBCO
nanoparticles which radius exceeds the critical one $R_{0}=15$~nm. 
The nanoparticle with a critical radius $R_{0}$ still contains 75$\pi \cdot
10^{3}$ unit cells of YBCO (the YBCO unit cell volume is $\sim 0.180$~nm$^{3}$). 
For the optimally doped regime, the enhancement of the
superconducting critical temperature for the particles with sizes $R=25$~nm
can be up to 6~K what is in agreement
with the experimental findings for YBa$_{2}$Cu$_{3}$O$_{7-\delta }$
nanoparticles in Ref.~\cite{menon}. 
Also we can not exclude the visualized 
enhancement $\Delta T_{c}\gtrsim $1K in YBa$_{2}$Cu$_{3}$O$_{7-\delta }$\
films in Ref.~\cite{mozhaev} due to detected spherical particle formation on
their surface in the process of laser ablation with increased deposition
temperatures $>790^{0}$~C. 
The developed theoretical approach can be applied to
nanoparticles prepared from other high-$T_{c}$ materials with narrow
electronic energy bands in bulk.

\bigskip

\bf{Acknowledgements}\rm{ -}
We acknowledge useful discussions with F.~Brosens. 
This work has been supported by GOA BOF UA 2000, IUAP, the FWO-V projects
Nos. G.0306.00, G.0274.01, WOG WO.025.99N (Belgium), and the ESF Programme
VORTEX.

\end{document}